\documentclass[fleqn,usenatbib]{mnras}

\usepackage{newtxtext,newtxmath}

\usepackage[T1]{fontenc}

\DeclareRobustCommand{\VAN}[3]{#2}
\let\VANthebibliography\thebibliography
\def\thebibliography{\DeclareRobustCommand{\VAN}[3]{##3}\VANthebibliography}

\usepackage{graphicx}	
\usepackage{amsmath}	

\title[Magnetar Polarimetry]{Probing magnetar emission mechanisms with spectropolarimetry}
\author[Caiazzo et al.]{
Ilaria Caiazzo\thanks{email: ilariac@caltech.edu; Sherman Fairchild Fellow}$^1$,
Denis Gonz\'alez-Caniulef$^2$\thanks{dgonzalez@phas.ubc.ca; CITA National Fellow},
Jeremy Heyl\thanks{email: heyl@phas.ubca.ca}$^2$
Rodrigo Fern\'andez$^3$
\\
$^{1}$TAPIR, Walter Burke Institute for Theoretical Physics, Mail Code 350-17, Caltech, Pasadena, CA 91125, USA\\
$^{2}$Department of Physics and Astronomy, University of British Columbia, Vancouver, BC V6T 1Z1, Canada\\
$^{3}$Department of Physics, University of Alberta, Edmonton, AB T6G 2E1, Canada.
}

\date{Accepted XXX. Received YYY; in original form ZZZ}

\pubyear{2015}

\begin{document}
\label{firstpage}
\pagerange{\pageref{firstpage}--\pageref{lastpage}}
\maketitle
\begin{abstract}
Over the next year, a new era of observations of compact objects in x-ray polarization will commence.  Among the key targets for the upcoming Imaging X-ray Polarimetry Explorer mission, will be the magnetars 4U~0142+61 and  1RXS~J170849.0-400910.  Here we present the first detailed predictions of the expected polarization from these sources that incorporate realistic models of emission physics at the surface (gaseous or condensed), the temperature distribution on the surface, general relativity, quantum electrodynamics and scattering in the magnetosphere and also account for the broadband spectral energy distribution of these sources from below 1~keV to nearly 100~keV.  We find that either atmospheres or condensed surfaces can account for the emission at a few~keV; in both cases either a small hot polar cap or scattering is required to account for the emission at 5-10~keV, and above 10~keV scattering by a hard population of electrons can account for the rising power in the hard X-rays observed in many magnetars in quiescence.  Although these different scenarios result in very similar spectral energy distributions, they generate dramatically different polarization signatures from 2-10~keV, which is the range of sensitivity of the Imaging X-ray Polarimetry Explorer. Observations of these sources in X-ray polarization will therefore probe the emission from magnetars in an essentially new way.
\end{abstract}



\section{Introduction} \label{sec:intro}

The study of neutron stars (NSs) has shown substantial progress in recent years thanks to new gravitational wave detectors, ground observatories, and space missions. The next generation of space telescopes able to  perform polarimetry observations will not only be a fantastic addition to the new era of multimessenger astronomy, but they will also open a new window to study a plethora of physical processes in different classes of NSs.  
In the soft X-ray band, the Imaging X-ray Polarimetry Explorer \citep[IXPE, NASA SMEX mission,][]{weisskopf16}, and the enhanced X-ray Timing and Polarimetry \citep[eXTP,  CNSA-ESA mission,][]{zhang16}, will operate in the  1$-$10 keV range and are schedule for launch in 2021 and 2027, respectively. A small CubeSat, PolarLight \citep{feng19}, recently measured the polarisation of the Crab nebula using the same gas pixel detector technology of many upcoming polarimeters \citep{feng20,long21}. On the other hand, while space-based polarimetry missions able to perform hard X-rays observations are still under study \citep[e.g. XPP,][]{jahoda19}, balloon-borne hard X-ray polarimeters like  PoGO+ \citep[$20-180$~keV,][]{chauvin16a} and X-Calibur \citep[$20-40$~keV,][]{gunther17}  have
already successfully tested hard X-ray polarimeters by performing  short observations of the Crab system \citep{chauvin16b} and the accretion powered pulsar GX 301-2 \citep{abarr20}, respectively.

Due to the rich phenomenology and extreme magnetic conditions,  Anomalous X-ray Pulsars (AXPs)  and Soft-Gamma Repeaters (SGRs) will be targets of great interest for the upcoming polarimetry missions. They belong to the class of strongly magnetized NSs, $B\sim10^{14}-10^{15}$~G, characterized by X-ray luminosities $L_X \sim 10^{33}-10^{36}\,\mathrm{erg~s}^{-1}$, substantially higher than their spin-down luminosity, and powerful hard X-ray or soft gamma-ray busting activity with luminosities ranging from $10^{39}$ to $10^{47}\,\mathrm{erg~s}^{-1}$ in timescales $\sim 0.1-40$~s. Their persistent spectra show generally both a soft X-ray component from 1 to 10 keV, well described by either multiple blackbodies or a blackbody plus power law, and a rising hard X-ray component extending from 10 up to 100 keV, well described by a single power law \citep[for a review see][]{turolla15,kaspi17}. The bulk of the AXP/SGR phenomenology has been successfully explained and unified within the magnetar model \citep{duncan92, thompson93,thompson95}. 
This considers AXPs and SGRs belonging the same class of NSs, called magnetars, whose decaying super-strong internal magnetic fields supplies both the heat source for the X-ray thermal spectrum and strong Lorentz forces acting on the NS surface. These forces can deform and even crack the NS solid crust, transferring a toroidal component from the NS interior to the external magnetic field, which can lead to the formation of a current-carrying, twisted magnetosphere. Within the magnetar model, the variety of bursting activity as well as the power laws observed in the broad X-ray energy band are though to be linked to this twisted magnetosphere.  

However, the emission mechanism powering the radiation in different parts of the spectrum of magnetars is not fully understood yet. The strong magnetic field present in these sources hints that, depending on the mechanism, the radiation might be substantially polarized. In fact, numerous studies have been dedicated to investigate the opacities and transport of radiation in strongly magnetized atmospheres, showing that the surface emission can be nearly fully polarized   \citep[for a review see e.g.][]{harding06,potekhin14}. In addition, the surface radiation and polarization can be further reprocessed by moving charged particles in the twisted magnetosphere, where resonant Compton scattering (RCS) can contribute a significant optical depth for X-ray photons \citep{thompson02}. The polarization evolution of light as it travels through the surroundings of the magnetar is also expected to be affected by the magnetized vacuum: according to quantum electrodynamics (QED), the vacuum can become birefringent \citep{heyl00,heyl02}. In order to investigate the above effects, Monte Carlo simulations have been developed to compute the spectra of magnetars \citep{fernandez07,nobili08}, showing that RCS by magnetospheric currents is able to explain the power-law observed in the hard X-ray emission from these sources. These works have been also expanded to study the polarization evolution in the magnetosphere \citep{fernandez11,taverna14} confirming the importance of future X-ray polarimetric observations both to test the QED birefringence of the vacuum and to disentangle the geometry of magnetars, as well as to probe whether they may have a gaseous or condensed surface  \citep[][see also \citealt{gonzalezcaniulef16} for an analogous study of X-ray dim isolated NSs]{taverna20}. 

The goal of this work is to revisit the problem of emission and transport of polarized radiation in magnetars. We model the X-ray persistent emission accounting for a varying surface temperature map, while considering emission either from an atmospheric surface or a condensed one, which allow us to include a missing ingredient in previous models: the emission from hot polar caps, more colloquially called hot spots. Furthermore, we account for RCS effects as well as alternative mechanisms that might explain the soft X-ray spectra of magnetars as is the case of Comptonization by a population of hot electrons just above the atmosphere. We perform numerical simulations for the various emission mechanisms considering the prototype magnetar 4U 0142$+$61 as a representative case.
The organization of the article is as follows. In Section 2, we describe the theoretical framework and physical ingredients  to  compute the emergent spectrum and polarization from a magnetars.
In Section 3, we show the results obtained of our simulations for 4U 0142$+$61 and apply them to the somewhat more energetic magnetar 1RXS~J170849.0-400910 as well. A summary of our main conclusions and discussion is given in Section 4.

\section{Theoretical framework}
\label{sec:Theoretical_framework}

The X-ray emission and polarization from magnetars are determined by the processes at the surface, magnetosphere, and magnetized vacuum around these sources. The persistent emission from magnetars in the soft X-rays peaks at about 1~keV and is usually interpreted as thermal emission from the neutron star surface. The thermal emission is thought to be reprocessed in the magnetosphere, because the soft emission can be well fitted with an absorbed blackbody with an excess above the peak (between 1 and 10 keV). The excess can be described by a steep power law with photon index $\sim2-4$, or by a second, hotter blackbody component.
The thermal component depends on atmospheric properties as well as on the temperature and magnetic map over the whole magnetar surface. The ``soft excess'', on the other hand, might be produced by a number of processes, such as a hot-spot or RCS or Comptonization, that to a varying degree are linked to the magnetospheric currents.
Many persistent sources also show a hard power-law, with a positive slope, that extends to hundreds of keV, for which proposed mechanisms range from thermal bremsstrahlung in the surface layers of the star, heated by a downward beam of charges, to emission from pairs created in the magnetosphere \citep{thompson05,thompson20}, to resonant Compton scattering (RCS) of seed photons on a population of highly relativistic charges \citep{baring05,fernandez07,nobili08,baring08,zane09,beloborodov13}. In this work, we will only consider RCS as the emission mechanism for the hard power law. Furthermore, the propagation of polarized radiation can be substantially affected by the magnetized vacuum, which can become birefringent under a strong magnetic field. In the following, we discuss separately each of these processes and their polarization pattern.

\subsection{Magnetized atmosphere}
\label{subsec:atmosphere}

In order to compute atmosphere spectra, several effects need to be taken into account.  In a magnetized plasma, radiation propagates in the so-called ordinary and extraordinary modes, which are commonly referred as O-mode and X-mode, respectively. These modes correspond to electromagnetic waves with the electric field direction oscillating in a direction either parallel (O-mode) or perpendicular (X-mode) to the plane defined by the propagation direction and the magnetic field. The opacities associated to each mode depend on the energy and propagation direction of the electromagnetic wave, as well as the strength and direction of the magnetic field in the plasma. For electromagnetic waves well below the electron cyclotron energy $E\ll E_\mathrm{c}= \hbar \mathrm{e} B/ m_\mathrm{e} \mathrm{c} ~\sim 1\,\mathrm{MeV} (B/10^{14}\mathrm{\,G})$, as the soft X-ray emission from magnetars, the opacities show large deviations with respect to the non-magnetic ones. In particular, for a direction of propagation (mostly) perpendicular to the magnetic field, the free-free and scattering opacities in the X-mode are strongly suppressed with the respect to the O-mode by a factor $\sim (E/E_\mathrm{c})^2$. As a consequence, the transport of radiation in the atmosphere becomes highly polarized and dominated by X-mode emission. On the other hand, for electromagnetic waves propagating in a direction (nearly) parallel to the magnetic field, both opacity modes become suppressed and, therefore, the transport of radiation becomes highly beamed along the magnetic field direction \citep[see e.g.][]{meszaros92,pavlov94,ho01}.   

The first self-consistent, magnetized atmosphere models investigated the propagation of radiation in a fully ionized, pure H plasma, with magnetic fields of the order $B\sim10^{12-13}$~G \citep{shibanov92,pavlov94}. Later, atmospheres with magnetic fields as strong as those of magnetars, $B>10^{13}$~G, were studied for angle-dependent scattering, using different numerical methods \citep[see e.g.][]{zane00,ho01,ozel01,lloyd03a}. Further works have also investigated and demonstrated the importance of several effects such as vacuum polarization \citep{zane01,ho03,vanadelsberg06}, ion cyclotron line creation \citep{zane01,vanadelsberg06}, partial ionization due to the magnetic field \citep{ho03a,potekhin04a}, different atmospheric compositions \citep{rajagopal97,mori07}, as well as a potential condensed surface or a He layer below the hydrogen atmosphere \citep{ho07,suleimanov09}. For a review, see e.g. \citealt{potekhin14a}. 

Atmosphere models for magnetars, in general, have considered these objects as passively cooling NS, i.e., they assume that the heat source is in the NS interior and the propagation of radiation in the atmospheric layers occurs without additional energy sources. However, the magnetosphere can sustain large currents of charged particles (see Sec. \ref{subsec:RCS}).  The magnetospheric particles can return to the surface with large Lorentz factor, depositing kinetic energy at different optical depths in the atmosphere. The transfer of radiation for the surface of a magnetar under a particle bombardment have been recently studied in the simplified case of a gray atmosphere \citep{gonzalezcaniulef19}, showing that the polarization pattern can be substantially reduced, or even become O-mode dominated (at variance of the X-mode radiation of passive atmospheres mentioned above). This work considered an uniform energy deposition in the atmosphere, representing a highly unknown momentum distribution of magnetospheric particles. However, for sufficiently high energy particles, their kinetic energy deposition can be located at very high optical depths in the atmosphere, producing spectra similar to passive cooling NS atmospheres. In the following, we consider this latter scenario, i.e, heat source well below the photosphere, using the model developed by Lloyd et al. \citep{lloyd03a,heyl03,heyl04}, whose numerical implementation  is discussed below. 

\subsubsection{Lloyd's Models}
Lloyd's method is very efficient at computing light-element (hydrogen and/or helium), plane-parallel atmospheres in radiative equilibrium in the limit of complete ionization, and is extensible to partial ionization \citep{lloyd03a}.  Also, in the code, the direction of the local magnetic field is allowed to vary from the vertical direction, and the effects of plasma birefringence are included self-consistently.  Although effects of vacuum polarization can also be included, we do not include these effects in the atmosphere. In the models, the atmosphere is assumed to be in hydrostatic equilibrium (any bulk motion is neglected). The pressure at any depth is the sum of ideal gas pressure, radiation pressure and non-ideal effects rising from Coulomb interactions in the ionized plasma. The ideal gas pressure includes the contribution from the degenerate pressure of electrons. In a strongly magnetized plasma though, electrons are forced into Landau levels, and the phase space volume occupied by the electron distribution is small; therefore, the onset of degeneracy occurs at higher densities compared to a weakly-magnetized plasma and the degeneracy pressure contributes for less than $\sim4\%$ even in the deepest layer. 

The principal opacity sources in the ionized plasma are Thomson scattering and free-free absorption. The presence of a strong magnetic field creates a preferred orientation to scattering and absorption processes, modifying the cross sections in the two modes and generating a finite polarization in the propagating radiation. Cyclotron resonances are treated by the self-consistent inclusion of ions and vacuum in the plasma dielectric.

The X-ray spectrum and polarization are found by iteratively solving the radiative transfer equations over a mesh in energy, polar angles and depth. The code assumes a plane-parallel atmosphere, which is a very good approximation since the atmosphere is incredibly thin compared to the radius of the star (centimeters compared to kilometers). However, the magnetic field varies in magnitude and direction across the surface of the neutron star, and therefore the surface of the star is divided in small patches in which the direction and strength of the magnetic field can be considered approximately constant. In order to get the total spectrum and polarization, one needs to sum over all the different patches.

\subsection{Emission from a condensed surface}
\label{subsec:condensed}
A strong magnetic field can change the energy levels of the atoms, and further confine the electrons in the direction perpendicular to the field.  The atoms become elongated along the magnetic field lines and can form molecular chains by covalent biding, which can lead via chain-chain interactions to the formation of a condensate \citep{ruderman71, lai97}. The phase transition from a gaseous to a condensed state depends on the temperature and composition of the NS surface, and heavy-element compositions are characterized by a higher critical temperature than light-element compositions. However, the properties of condensed surfaces remain highly unknown. Numerical results for the calculations of critical temperatures can be described  by  $T_\mathrm{crit} \approx 5\times10^4 Z^{1/4} (B/10^{12}\,\mathrm{G})^{3/4}$K in the magnetic field range $10^{12}\mathrm{G}\la B\la 10^{15}$G \citep{lai01,medin07,potekhin16}. A plausible scenario is, therefore, that magnetars with sufficiently strong magnetic fields and mild surface temperatures might be emitting thermal radiation directly from a metallic or liquid surface. 

A number of works have investigated and modelled the spectra from condensed surfaces \citep{brinkmann80,turolla04,perezazorin05,vanadelsberg05,potekhin12}. The emergent spectrum depends on the emissivity of the surface, which in turn depends on the reflectivity through  Kirchhoff’s law. However, the response of the ion lattice to electromagnetic waves in the condensate is still unknown. The dielectric tensor for the condensed surface is usually considered in two limiting cases: i) {\it fixed ions}, meaning that the ions do not have any response to electromagnetic waves, as Coulomb forces in the lattice between ions are dominant, or ii) {\it free ions}, meaning that the ions respond fully to electromagnetic waves.  Models for the polarized emission from strongly magnetized sources with condensed surfaces have been presented by \citet{gonzalezcaniulef16} for X-ray dim isolated neutron stars, and similarly by \citet{taverna20} for the case of magnetars.

In the following, we compute the emissivities from condensed surfaces using the analytical approximations of \citet{potekhin12}. They are defined in modes 1 and mode 2, which in turn are defined relative to the normal to the surface and propagation direction. The transformation to the X-mode and O-mode is given in the appendix B of \citet{potekhin12}. The intensity of the radiation for these modes are:
\begin{eqnarray}
I_o &=& j_o(\nu, B, k,\theta_{Bk})\, B_{\nu}(T)\\
I_x &=& j_x(\nu, B, k,\theta_{Bk})\, B_{\nu}(T),
\end{eqnarray}
where $B$ is the strength of the magnetic field, $k$ is propagation direction of the electromagnetic wave, $\theta_{Bk}$ is the angle between the local magnetic field and propagation direction, and  $B_{\nu}(T)$ is the blackbody function. Additional details and discussions about the numerical implementation of condensed surface models can be found in \citet{gonzalezcaniulef16}.  We restrict the modelling of condensed surfaces to iron composition. For simplicity we also consider only the case of a dielectric tensor in the limit of fixed-ions, whose spectrum in the soft X-rays is slightly more polarized than for the case of free-ions (although in general, condensed surfaces produce low degrees of polarization of a only few per cent). Further discussions of polarized emission in the limit of free-ions can be found in \citet{gonzalezcaniulef16} and \citet{taverna20}. 

\subsection{Temperature map and hot spots}
\label{subsec:Temperature_map}

The effective temperature over the surface of a magnetar varies as result of the crustal conductivity and non-uniform topology of the strong magnetic field, which restrict the electrons to move just along the magnetic field lines \citep[see e.g.][]{greenstein83,heyl98,potekhin15b}.  In order to calculate the thermal emission, we divide the magnetar surface in several patches. Specifically, for a patch at co-latitude $\theta$ with respect to the magnetic axis, the local magnetic field strength is given by (assuming a dipolar field):
\begin{equation}
    B = B_p \sqrt{ \frac{3\cos^2\theta + 1}{4}}\,,
    \label{eq:BBp}
\end{equation}
where $B_p$ is the magnetic field at the pole. The angle between the local vertical and the magnetic field is given by
\begin{equation}
    \cos^2\psi = \frac{4\cos^2\theta}{3\cos^2\theta +1}\,;
\end{equation}
while the flux is
\begin{equation}
    F = F_p \left(\frac{B}{B_p}\right)^{0.4}\cos^2\psi = F_p  \left[\sqrt{ \frac{3\cos^2\theta + 1}{4}}\right]^{0.4}\frac{4\cos^2\theta}{3\cos^2\theta +1}\,,
\end{equation}
where $F_p$ is the flux at the pole \citep{heyl98}. The effective temperature of the patch is therefore
\begin{equation}
    T_{\rm{eff}} = T_{\textrm{eff}p}  \left[\left( \frac{3\cos^2\theta + 1}{4}\right)^{0.2}\right]^{1/4}\left(\frac{4\cos^2\theta}{3\cos^2\theta +1}\right)^{1/4}\,,
    \label{eq:Tdep}
\end{equation}
where $T_{\textrm{eff}p} $ is the effective temperature at the pole.  In contrast to \citet{2021arXiv211108584K}, we choose the effective temperature at the pole that can reproduce the observed spectral energy distribution and we impose for the flux to vary across the surface following a dipole distribution.

\begin{figure}
    \centering
    \includegraphics[width=0.47\columnwidth]{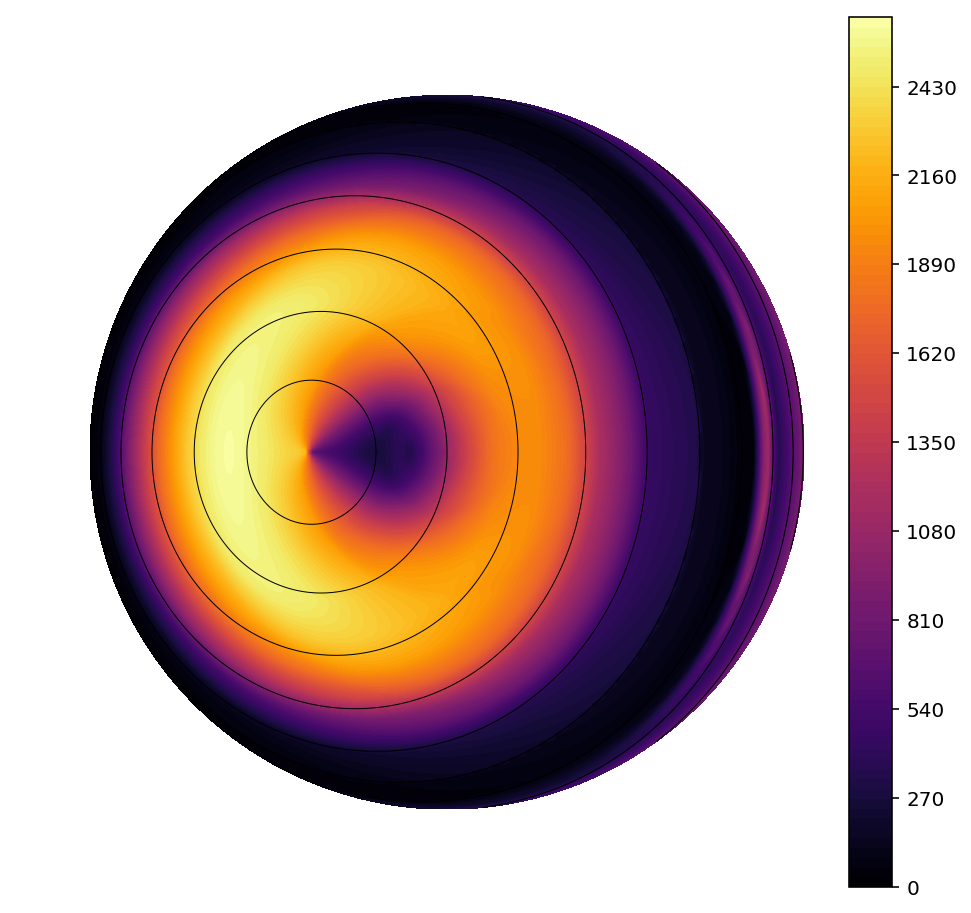}
    \includegraphics[width=0.47\columnwidth]{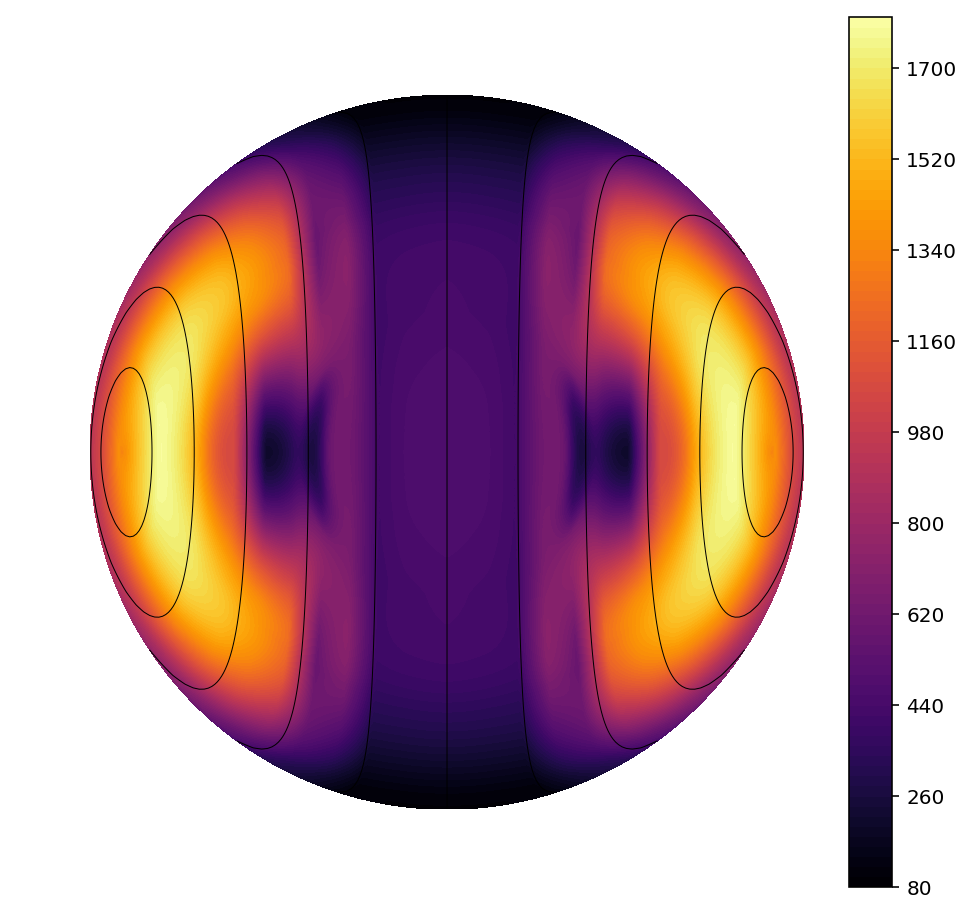}
    \caption[Intensity map for the thermal emission.]{Intensity map for the thermal emission of a magnetar with $T_{\textrm{eff}p}=3.0\times10^6$~K and $B_p=1.3\times10^{14}$~G, including light bending. Left: viewing angle $30^\circ$; right: viewing angle $90^\circ$. Black circles indicate contours of equal colatitude, The colormap shows the intensity of the emission for 2~keV photons in arbitrary units.}
    \label{fig:MapSurf}
\end{figure}

As an example, Fig.~\ref{fig:MapSurf} shows the intensity map for the atmospheric emission (at 2~keV) from a magnetar with the magnetic pole at $30^\circ$ (left) and at $90^\circ$ (right) from the line of sight. The black circles are drawn at constant co-latitudes. Gravitational light bending is included in the calculation, and therefore both magnetic poles are visible in the $90^\circ$ case. To calculate the intensity, we divided the surface in 18 patches of equal luminosity (from one pole to the other), and for each patch we calculated the atmosphere emission using Lloyd's code. The effective temperature and magnetic field at the pole are taken to be $T_{\textrm{eff}p}=2.9\times10^6$~K and $B_p=1.3\times10^{14}$~G, and the temperature and magnetic field for each patch are calculated using Eqs.~\ref{eq:BBp} and~\ref{eq:Tdep}. The $\cos^2 \psi$ dependence can be seen in the map. The dark region close to the pole in the left image and the two dark regions at mirrored positions with respect to the center in the right image correspond to the regions where the local magnetic field is pointing in the direction of the line of sight. This seems counter-intuitive, as photons can escape more easily when streaming along the magnetic field line; however, we cannot resolve the very small region in angle around the magnetic field direction where the intensity peaks, while the region around the peak is depleted of photons because they can easily get scattered in the magnetic field direction.

The spectrum of several magnetars in quiescence can be fitted with a double black body \citep[see for example][]{2015ApJ...808...32T}, which hints to the possibility of the excess above the thermal peak to be caused by a hot spot. 
In order to model the surface thermal emission, we also consider the emission from hot spots with uniform temperature at the magnetic poles. The temperature and size of the hot spot (in terms of the co-latitude angle) are set according the surface emission model i.e., atmosphere or condensed surface.   

\subsection{Compton scattering above the atmosphere}
\label{subsec:Compton}

An alternative explanation for the excess above the thermal peak, that has not been proposed before, is saturated Comptonization of the thermal photons by a non-relativistic population of electrons close to the stellar surface. This is a somewhat conservative model, as it does not require relativistic or ultra-relativistic electrons, or a specific distance from the star. However, because the majority of thermal photons are being emitted in the X-mode, and the scattering cross section for X-mode photons is very small, resonant scattering has been invoked in order to build enough optical depth to explain the power-law excess with Compton scattering of X-mode photons. This is not necessary though in the case in which at least a small fraction of thermal photons is emitted in the O-mode; from Lloyd's atmosphere models, the fraction of photons emitted in the O-mode is of the order of 2\%.

As aforementioned, the difference in scattering cross section between the two modes at these energies is several orders of magnitude, and therefore, for O-mode photons, it is much easier to build enough optical depth to fully Comptonize the population. Moreover, whenever an X-mode photon happens to scatter, it is immediately converted into an O-mode photon, because we are considering energies far below the electron cyclotron line, and after the scattering event the photon's scattering cross section is therefore hugely increased. 

The evolution of the spectrum in the presence of repeated scatterings off non-relativistic or mildly relativistic electrons can be calculated with the Kompaneets equation \citep{1975JETP...40..208K} and in case of saturated scattering, the equation leads to an approximated Wien law. This is because, when photons undergo many scatterings, they reach a thermal equilibrium with the electrons, and get ``scattered up'' into a Bose-Einstein distribution \citep{1986rpa..book.....R}. Even if only 2\% photons are in the O-mode, this is enough to explain the observed excess above the thermal peak if the scattering plasma has a temperature of $\sim2$~keV.

\begin{figure}
    \centering
    \includegraphics[width=0.465\columnwidth]{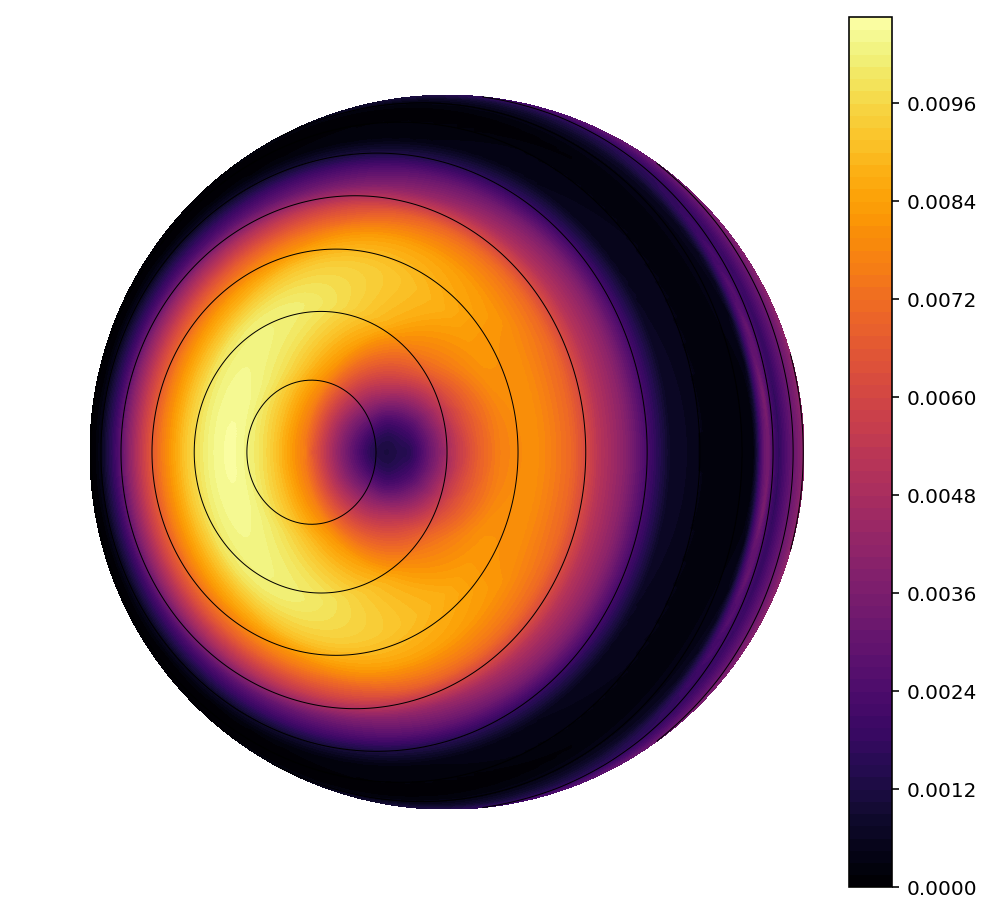}
    \includegraphics[width=0.465\columnwidth]{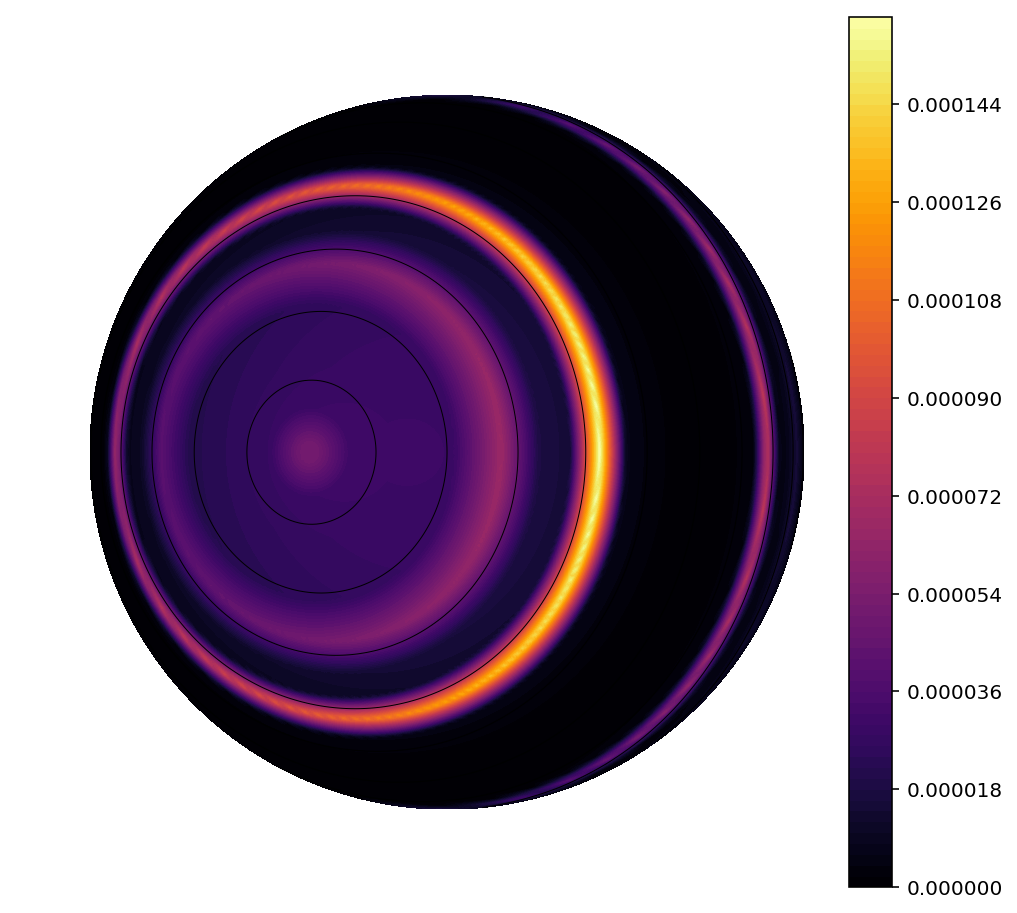}
    \caption{Intensity map for the saturated Comptonization model, for a viewing angle $30^\circ$. The left image shows the intensity for the X-mode photons, while the right image is intensity of the O-mode photons, both at 5~keV. The total intensity looks like the image to the left for energies below 6-7 keV; for higher energies, the O-mode photons begin to dominate and the total intensity starts looking like the right image.}
    \label{fig:MAPCompO}
\end{figure}

In order to calculate the emission spectrum from the magnetar in the case of full Comptonization of O-mode photons, we therefore take the thermal emission from the surface and divide the photons into X-mode and O-mode. The X-mode photons escape freely and preserve the original distribution, while for the O-mode photons we calculate the Comptonized spectrum assuming that they reach a Bose-Einstein distribution. For the surface thermal emission, we assume a pole temperature of $kT_{\textrm{eff}p} = 0.32$~keV and a magnetic field at the pole $B_p = 1.3\times10^{14}$~G, and we compute the atmosphere models at the surface, following the equations in \S~\ref{subsec:Temperature_map}. For each patch, the O-mode radiation gets scattered up by an electron-dominated plasma at a temperature of $kT_e = 2.1$~keV, and reaches a Bose-Einstein distribution with a chemical potential $ \mu/kT_e\sim10$. We calculate the chemical potential by making sure that the total number of photons remain the same before and after scattering. Fig.~\ref{fig:MAPCompO} shows the intensity map for the X-mode and the O-mode on the surface of the neutron star for a viewing angle of $30^\circ$. The left panel shows the intensity map for the X-mode photons at 2~keV, which escape directly from the atmosphere without scattering. As most of the atmosphere photons are in the X-mode, the intensity shown in the left panel is very similar to the total thermal emission shown in Fig.~\ref{fig:MapSurf}. The right panel shows the intensity map for the O-mode photons also at 2~keV, which scatter many times in a hot corona right above the neutron star surface. Because of the inclination of the magnetic field with respect to the surface, the atmosphere patch at $45^\circ$ produces a higher fraction of O-mode photons, and a bright ring is shown at about $45^\circ$ in the right panel.

\subsection{Vacuum birefringence in the magnetosphere}
\label{subsec:Vacuum_birefringence}

The presence of a strong magnetic field can make a medium birefringent: the index of refraction in the medium depends on the angle between the polarization of the photon and the magnetic field \citep{heisenberg36,weisskopf36,1997JPhA...30.6485H}. In the case of a magnetized vacuum, the birefringence is caused by the interaction of photons with virtual electron-positron pairs: it is easier to excite virtual electrons along the direction parallel to the magnetic field than perpendicular to it, and thus photons in the ordinary mode travel slower than photons in the extraordinary mode. In a birefringent medium, in which the anisotropy is set by the magnetic field, the two polarization modes, parallel and perpendicular to the magnetic field, are decoupled if \citep{heyl00}
\begin{equation}
\left | \hat \Omega \left ( \frac{ \rm d \ln |\hat \Omega|}{ \rm d \lambda} \right )^{-1} \right | \geq 0.5\,. 
\label{eq:rp1}
\end{equation}
where $\lambda$ measures the length of the photon path in the medium and $\hat\Omega$ is the birefringence vector, given by
\begin{equation}
|\hat \Omega|=|k_0\Delta n| = \frac{\alpha}{15} \frac{\nu}{c} \left ( \frac{B_\perp}{B_\mathrm{\scriptsize QED}} \right )^2  \,.
\label{eq:omegahat}
\end{equation}
In this case, the evolution is called adiabatic, and the photon polarization follows the direction of the local field lines.

In the case of neutron stars, assuming a dipolar magnetic field ($B \approx \mu r^{-3}$, where $\mu$ is the magnetic dipole moment of the star and $r$ is the distance from the center of the star) the adiabatic condition of eq.~\ref{eq:rp1} translates into \citep{heyl02}
\begin{equation}
\left | \frac{\alpha}{15} \frac{\nu}{c} \frac{\mu^2 \sin^2 \beta}{r^6 B^2_\mathrm{\scriptsize QED}} \frac{r}{6}  \right | \geq 0.5
\label{eq:rp2}
\end{equation}
where $\beta$ is the angle between the dipole axis and the line of sight. If we define the polarization-limiting radius ($r_\mathrm{\scriptsize PL}$) to be the distance at which the equality holds, we find that the polarization will follow the direction of magnetic field out to
\begin{align}
r_\mathrm{\scriptsize PL} &= \left ( \frac{\alpha}{45}
 \frac{\nu}{c} \right )^{1/5} \left ( \frac{\mu}{B_\mathrm{\scriptsize QED}} \sin
 \beta \right )^{2/5} \nonumber \\
 &\approx  1.2 \times 10^{7} \left
 ( \frac{\mu}{10^{30}~\mathrm{G~cm}^3} \right )^{2/5} \left (
 \frac{\nu}{10^{17}~\mathrm{Hz}} \right)^{1/5} \left ( \sin \beta
 \right)^{2/5} \mathrm{cm.}
 \label{eq:rpl}
\end{align}

For X-ray photons coming from near the surface of a neutron star with a surface field of $10^{14}$~G, the polarization-limiting radius is much larger than the star, according to eq.~\ref{eq:rpl}, so the observed polarization of the photons will reflect the direction of the magnetic field at a large distance from the star and not at the surface.  For a much more weakly magnetized star, the polarization-limiting radius will be comparable to the radius of the star, so the observed polarization will reflect the field structure close to the star.

If QED is not included, the polarization observed at infinity is the sum of the contribution from the whole surface, where the magnetic field is pointed in many different directions and so is the polarization. Even if the surface emission is 100\% polarized, the total observed polarization is very low, just a few percent. The effect of QED in presence of a high magnetic field is to preserve the polarization degree at emission: if one includes QED, the polarization direction is not frozen to the value at the surface, but keeps changing following the local magnetic field to the polarization-limiting radius, tens of stellar radii away from the surface in the case of magnetars, where the magnetic field through which the radiation passes is uniform. If a photon is emitted in the X-mode, its polarization will rotate so that it keeps being perpendicular to the local magnetic field (the photons keeps staying in the X-mode) and the same for an O-mode photon \citep{heyl02,heyl03,heyl04}.

\subsection{Resonant Compton scattering in the magnetosphere}
\label{subsec:RCS}

RCS effects by currents of charged particles can contribute a significant optical depth for photons propagating in the magnetosphere of magnetars \citep{thompson02}. RCS can boost surface thermal photons to higher energies and give rise to a power-law as such observed in the X-ray spectrum of magnetars  \citep{fernandez07,nobili08,baring08,zane09}. The formation of such a power-law spectrum depends on the configuration of the non-potential magnetosphere, which sets the currents of charged particles i.e.,  $\nabla \times \mathbf{B} =  4\pi \mathbf{j} /c$. 
For self-similar, twisted magnetospheres, solutions are obtained by solving the Grad-Shafrenov equation. In particular, for an axisymmetric dipolar field, the magnetic field components in spherical coordinates are  given by 
\begin{equation}
 \mathbf{B}= \frac{B_{\mathrm{p}}}{2} \left(\frac{r}{R_\textrm{NS}}\right)^{-p-2}
  \left[-f'(\mu), \frac{pf(\mu)}{\sin\theta},
    \sqrt{\frac{C(p)\ p}{p+1}} \frac{f^{1+\frac{1}{p}}(\mu)}{\sin\theta}\right]
\end{equation}
where the function $f(\mu)$  depends on $\mu = \cos \theta$, and the input parameter $p$ specifies the magnetospheric shear \citep{thompson02} . Alternatively, the shear can also be specified by the twist angle
\begin{eqnarray}\label{twistangle}
  \Delta \Phi &=&
  \int_{\textrm{field line}} \frac{B_\phi}{(1-\mu^2) B_\theta} \mu d\mu
  \nonumber \\
  &=& \left[\frac{C(p)}{p\ (1+p)}\right]^{1/2}
  \int_{\textrm{field line}} \frac{f^{\frac{1}{p}}(\mu)}{1-\mu^2} \mu d\mu \, ,
\end{eqnarray}
which can vary from a pure dipolar field 
($\Delta \Phi =0$, or $p =1$) to a split monopole configuration ($\Delta \Phi =\pi$, or $p=0$). 

By setting the twist angle, the magnetic field configuration is determined,  as well as the magnitude of the magnetospheric currents. However, in order to compute the transport of radiation, the particle flow needs to be further specified by the type of charge carriers and their momentum distribution. More explicitly, the magnetospheric  current is given by
\begin{equation}
\mathbf{J} = \sum_i Z_ie\,n_i\, \bar\beta_i\hat{B}
c = \frac{(p+1)c}{4\pi r}\frac{B_\phi}{B_\theta}\, \mathbf{B},
\end{equation}
where $Z_ie$ is the electric charge of species $i$,
$n_i$ the number density, and $\hat B = \mathbf{B}/B$. Here, the  mean velocity 
\begin{equation}
\bar\beta_i = \int f_i(\mathbf{r},p)\, \frac{pc}{E}\, d^3p
\end{equation}
can be satisfied by a variety of  phase-space distribution functions $f(\mathbf{r},p)$. Typically, the formation of the non-thermal component of magnetar spectra has been studied considering an electron-ion  or electron-positron plasma, whose momenta follow a single  power-law  or thermal distribution \citep[see e.g.,][]{fernandez07,nobili08}, showing that reprocessing of surface radiation by RCS in general drives the overall polarization fraction from magnetars toward about 30\% \citep{fernandez11,taverna14,taverna20}. 

In the following, the transport of radiation in the magnetosphere is solved using the Monte Carlo code discussed in \citet{fernandez07} and \citet{fernandez11}. As in the original version of the code, we consider an electron flow with a single power-law momentum distribution $f(\gamma\beta) \propto (\gamma\beta)^{-\alpha}$, further characterized by  the minimum
velocity $\beta_{\mathrm{min}}$, and the maximum Lorentz factor $\gamma_{\mathrm{max}}$.  Alternatively, we also expanded the code to account for electrons with a broken power-law momentum distribution given by
\begin{equation}
f(\gamma\beta) \propto \left\{
        \begin{array}{ll}
            (\gamma\beta)^{-\alpha_1} & \quad \gamma\beta < (\gamma\beta)_\mathrm{break} \\
            (\gamma\beta)^{-\alpha_2} & \quad \gamma\beta \geq (\gamma\beta)_\mathrm{break},
        \end{array}
    \right.
\end{equation}
where the momentum  $(\gamma\beta)_\mathrm{break}$ is an additional free parameter. The choice for this electron momentum distribution is motivated by the observation of 4U 0142+61 and enable us to describe simultaneously both soft and hard X-ray power-law components in the spectrum of this magnetar (see Section \ref{subsec:mag_4U}). 

Previously, the code considered only the case of blackbody emission with uniform temperature either over polar caps or the whole surface, but it has been expanded to account for the magnetar surface emission with a varying magnetic field and thermal map discussed above (either atmospheric or condensed) and for the soft-excess emission (as either a hot spot or a Comptonized O-mode excess). In order to implement different surface models, we work under the assumption that the RCS scattering radius is located far from the NS surface, at several NS radii, far enough that the surface of the magnetar appears as a point source. This allows us to consider surface seed photons with radial trajectories. Then, for each radial trajectory,  the spectrum is simply described by the surface mean intensity at the corresponding magnetic co-latitude angle. Using this approach, the mean intensity can be pre-calculated with a ray tracing code that accounts for  i) a realistic surface model emission, ii) gravitational redshift and light bending due to general relativity, and iii) vacuum birefringence inside the RCS scattering radius.

Once the surface model is set, single seed photons are generated with an inverse transform sampling method. The assumption of radial trajectories translate to the the problem of sampling photons from a two-dimensional discrete distribution, i.e. mean intensity that depends on energy and co-latitude angle with respect to the magnetic axis. The sampling method can be summarized in the following steps:
\begin{enumerate}
 \item Compute the energy dependent cumulative distribution function (CDF)  after marginalizing the distribution for the mean intensity in the colatitude angle.
\item Generate a random photon with energy $\epsilon$ from the energy dependent CDF.
\item Derive the colatitude angle dependent mean intensity for the random energy $\epsilon$ and compute the associated CDF.
\item Generate a random co-latitude angle for the photon from the co-latitude angle dependent CDF.
\end{enumerate}
After setting the energy and co-latitude angle, the photon polarization is  generated from a uniform random distribution with either  probability $p = (1-Q)/2$ for the X-mode or the complement probability for the O-mode. Here $Q$ is the normalized Stokes parameter from the input surface model relative to the magnetic axis, while $U=0$ in this frame. Detailed explanation of the RCS code for the propagation of radiation in the magnetosphere with scattering and further evolution of the Stokes parameters with vacuum birefringence can be found in \cite{fernandez07} and \cite{fernandez11}. A key assumption of the treatment is that the energy of the photon in the frame of the electron is small compared to the rest-mass energy of the electron so that electron recoil can be neglected.  Because the seed photons typically have energies of 2$-$3~keV and the final photons energies that we consider are up to 200~keV, this is a reasonable assumption if the photons suffer typically at most one scattering \citep[c.f.][]{2008MNRAS.389..989N}.

\section{Simulations and Observational Prospects}
\label{sec:target_simulation}

In the following subsections we will describe how the various emission and scattering processes outlined above can account for the spectral energy distribution of the bright AXP 4U~0142+61 and the polarization signatures that result from these processes.  Furthermore, we will outline how these same processes can account for the observations of 1XRS J170849.0-400910 as well.

\subsection{4U 0142+61}
\label{subsec:mag_4U}

One of the prototypical persistent magnetars is 4U 0142+61. The soft X-rays can be described by a single blackbody component plus a power law, or a double blackbody, whereas the hard X-rays show an additional raising power law with index of $\Gamma_H \approx 0.8$ \citep{2015ApJ...808...32T}. Observations with RXTE show pulsations with a period of 8.7~s and spin-down rate $\dot P=2.02\times 10^{-12}$ \citep{2014ApJ...784...37D}. This translates into a spin-derived magnetic field of $1.3\times 10^{14}$~G.

\subsubsection{Atmospheric spectrum}
\label{subsubsec:atm_spec}
We model 4U 0142+61 considering three different atmospheric surface input models that either partially or fully describe the soft X-ray emission from this source:
\begin{enumerate}
\item Atmosphere with varying surface thermal map and magnetic field. The temperature at the magnetic pole is set in such a way that the atmospheric emission reproduces the low-energy part of the spectrum of 4U 0142+61, below $3$~keV.
\item Atmosphere plus hot spot.  We added hot spots to the varying surface thermal map, described in (i), in each magnetic pole. The temperature and area of the hot spot is chosen in such a way that the spectrum reproduces the whole soft X-ray spectrum ($2-10$ keV) of 4U 0142+61.
\item Atmosphere plus O-mode Comptonization. We consider a population of hot electrons (whose properties in this work are considered independent of the electron flow due to the twisted field) near the top of the atmosphere, that Comptonize the O-mode atmospheric photons, whereas the X-mode photons freely escape. The temperature of the hot electrons is set in such a way that overall emission reproduces the whole soft X-ray emission from 4U 0142+61, in the $2-10$~keV range .
\end{enumerate}

We perform the RCS Monte Carlo calculation \citep[][]{fernandez07,fernandez11} for the different surface emission models and velocity distributions of the magnetospheric particles.  We restrict to an unidirectional flow of electrons with either a single or broken power-law (PL) momentum distribution (see sec~\ref{subsec:RCS}). The magnetospheric twist angle, $\Delta \phi$, as well as the parameters that set the electron PL momentum distributions, $\beta_{\mathrm{min}}$, $\beta_{\mathrm{broken}}$, and $\gamma_{\mathrm{max}}$, are chosen in such a way that the RCS processed radiation approximately describes either the hard X-ray spectrum alone or the whole soft and hard X-ray spectrum of 4U 0142+61.  For simplicity, we model only the phase-averaged spectrum for this source, while the phase-dependent analysis is left for future work. In addition, we limit our analysis to the case of an orthogonal rotator, i.e. both the line-of-sight and magnetic axis form a 90 deg angle with respect to the spin axis of the magnetar. Although the emission geometry can be constrained to several degrees using the existing phase-resolved spectral energy distributions, the future polarimetric measurements will result in constraints many times stronger \citep{2021arXiv211108584K,gonzalezcaniulef_unbinned}, so we leave the precise geometry to be determined with the polarization data. All RCS simulations are run considering $10^7$ photons.  

Figure~\ref{fig:mc_atm} shows our simulations for RCS considering the different atmospheric surface emission models.  We attempt to simultaneously reproduce the soft and hard X-ray emission. This can be achieved either by adding hot spots to the surface thermal map (case ii) or by Comptonizing the atmospheric O-mode photons by a population of hot electrons near the top of the atmosphere (case iii) to reproduce the soft excess below 10 keV. In both cases, we need a relatively small magnetospheric twist $\Delta \phi =0.1$~rad and a hard power law for the electron momentum distribution with maximum Lorentz factor $\gamma_{\mathrm{max}}=30$, to reproduce the power law above 10~keV with RCS. Thus, the energetic electrons at several NS radii can transfer enough energy from the RCS process to the seed photons to form the hard X-ray emission of 4U 0142+61. The intensity and polarization signal of case (ii) plus RCS and case (iii) plus RCS are depicted in Figure~\ref{fig:mc_atm} as green and red lines respectively.
 
Alternatively, we explore the possibility of explaining both non-thermal components of 4U~0142+61, observed in soft and hard X-rays, considering RCS by an electron flow in the magnetosphere with a broken PL momentum distribution. For simplicity, we consider a soft PL component with index $\alpha_1 = 3$ and a hard PL component with index $\alpha_2=1$ (as the models discussed previously). The velocity transition from the soft to the hard electron PL component is set as a free parameter $\beta_{\mathrm{break}}$ that can be adjusted to the data. We find that employing just the atmosphere seed photons (case i), a $\beta_{\mathrm{break}}=0.9$ and a relatively large twist $\Delta \phi=0.6$~rad enable us to partially reproduce the whole X-ray spectrum of 4U~0142+61 (purple line in Figure~\ref{fig:mc_atm}). The evident difficulty to reproduce the spectrum in the  $\sim 10-11$ keV  range is due to the fact that the large magnetospheric twist translates into a large optical depth for RCS, leading to multiple scattering in this energy range of the spectrum (photons scatter about three times before leaving the magnetosphere).  In principle, many of the photons with final energies around 10~keV are down scattered and therefore may have had energies in the electron rest frame approaching or exceeding $m_e c^2$, so electron recoil could be important here \citep{2008MNRAS.389..989N}.  A more thorough exploration of the parameter space for RCS and a potential relaxation of the $\alpha_1 = 3$ and $\alpha_2=1$ conditions, as well as the inclusion of the effects of recoil, might enable us to produce a closer spectrum to 4U~0142+61 with resonant scattering alone.  In fact, outside of this regime for this particular RCS model (the broken power-law), the typical number of scatterings is zero or one, so neglecting electron recoil is appropriate for the other models.

Finally, we consider the case in which RCS might affect only the soft X-ray spectrum of 4U 0142+61, while the hard X-ray might be produced by a different mechanism. We find that with a twist angle  $\Delta \phi =0.4$~rad and a relatively low  $\gamma_{\mathrm{max}}=1.5$ for the electron momentum distribution, we can easily reproduce the soft X-ray spectrum of this source (orange line in Figure~\ref{fig:mc_atm}). 

The models presented above show, for the first time,  that it is possible to reproduce the broadband soft and hard X-ray spectrum of 4U~0142+61 considering different atmospheric emission mechanisms and the right tuning of the parameter space for RCS. The strongest condition is that the magnetospheric electron flow should be described by a broad PL momentum distribution (either single or broken PL). As shown in the early results of the RCS code used in our analysis \citep{fernandez07}, a Boltzmann particle distribution is not able to explain the slope of the hard X-ray emission of 4U 0142+61.  In order to distinguish between different emission mechanisms observationally, we need to understand their underlying polarization information, which is discussed below. 

\subsubsection{Atmospheric polarization}

In the Monte Carlo code, all models consider transport of radiation in the magnetosphere under vacuum birefringence. The polarization axes for a distant observer are defined relative to the spin axis of the magnetar. For example, for an aligned rotator, the electric field of X-mode dominated radiation oscillates perpendicular to the spin axis. On the other hand, for an orthogonal rotator, as the geometry considered in our simulations, the X-mode dominated radiation is parallel to the spin axis. To keep track of this information, positive Stokes $Q$ from the Monte Carlo output reflects X-mode dominated radiation, whereas negative Stokes $Q$ is associated to O-mode dominated radiation. 

\begin{figure}
\includegraphics[width=\columnwidth]{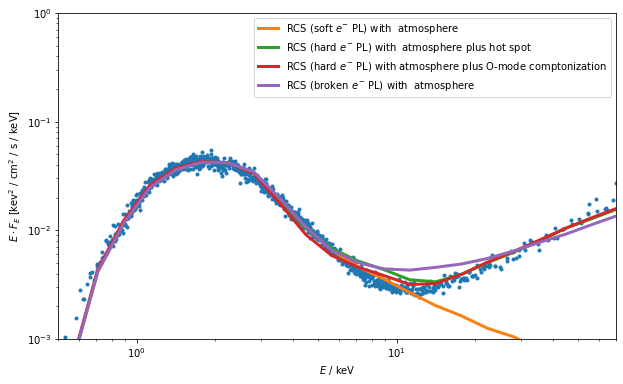}
\includegraphics[width=\columnwidth]{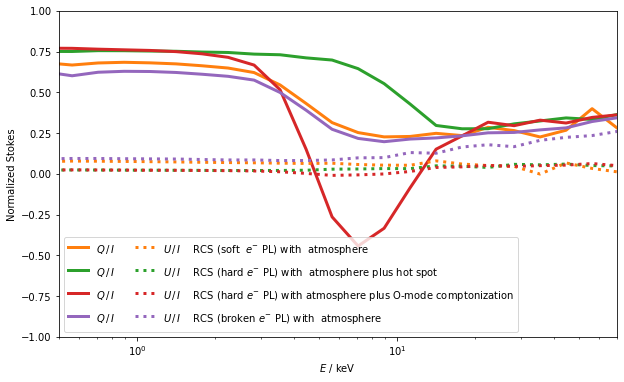}
\caption{Resonant Compton scattering (RCS) with different electron power law ($e^-$ PL) momentum distribution and atmospheric input surface models. See Sec \ref{subsubsec:atm_spec} and Table \ref{tab:1} for more details about the models. The points depict the unfolded spectral energy distribution from \citet{2015ApJ...808...32T}.
}
\label{fig:mc_atm}
\end{figure}

The lower panel of Figure~\ref{fig:mc_atm} shows the polarization signal for the different emission mechanisms that assume an atmospheric surface.  In the hard X-rays, all models converge to the standard polarization fraction $\sqrt{Q^2+U^2}/I\sim30\%$ expected from the different cross sections for RCS associated with mode exchange. In the soft X-rays, the emission and scattering mechanisms show substantially different polarization patterns. On one hand, the atmospheric emission including a hot cap results in a relatively large polarization fraction, well above $\sim60\%$, due to the dominance of X-mode radiation from the atmospheric magnetized plasma (green line). If one considers a population of hot electrons sufficient to Comptonize the atmospheric emission fully (red line), the result is a mode switch around $5$~keV. This is due to the fact that the population of hot electrons can up-scatter only the atmospheric O-mode photons, whereas the X-mode photons freely escape as their Thompson cross section is strongly reduced by the magnetic field. Lastly, if we try to explain the soft X-ray emission with RCS either using a single (orange line) or broken (purple line) PL momentum distribution for the electron flow, the onset of the typical $30\%$ RCS polarization fraction moves inside the soft X-ray band, just above $\sim 5$~keV. For all models, there is a non-null Stokes $U$ component whose magnitude increases for larger magnetospheric twist angles, which is quite evident in the hard X-rays.

\begin{figure}
\includegraphics[width=\columnwidth]{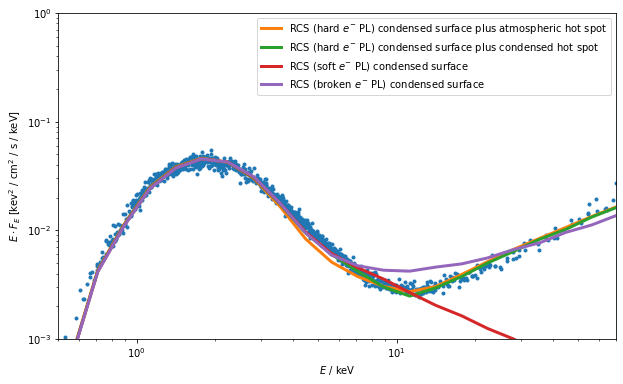}
\includegraphics[width=\columnwidth]{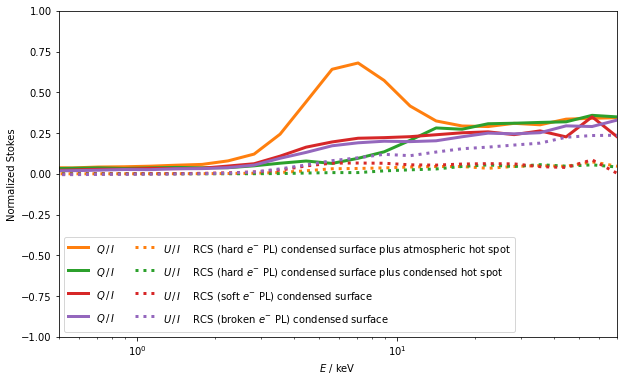}
\caption{RCS simulations similar to Figure \ref{fig:mc_atm}, but for condensed surface emission. Here, we model a surface map with either atmospheric or condensed hot spots. Full Comptonization is not considered. See Sec \ref{subsubsec:cond_spec} and Table \ref{tab:1} for more details.}
\label{fig:mc_cnd}
\end{figure}

\subsubsection{Condensed surface spectrum}
\label{subsubsec:cond_spec}
We now repeat our analysis of the spectrum of 4U 0142+61, but considering  thermal emission from a condensed surface (in the limit of fixed ions, see sec.~\ref{subsec:condensed}).  We model the following scenarios:
\begin{enumerate}
\item  Condensed surface with varying surface map and magnetic field. The surface model is set to reproduce the spectrum of 4U 0142+61 below $\sim 3$~keV.
\item Condensed surface with condensed hot spots. We add to the surface map (i) a hot spot at each magnetic pole with the temperature and size that enables us to reproduce the whole soft X-ray emission from  4U 0142+61.
\item Condensed surface with atmospheric hot spot. As in (ii), we add in each magnetic pole a hot spot which is now atmospheric, whose size needs to be relatively large to explain the whole soft X-ray emission from the source.
\end{enumerate}
Using similar RCS settings as those already simulated in the case of the atmospheric emission (magnetospheric twist and momentum distribution for the electron flow, see Table \ref{tab:1}), we also obtain similar results (see Figure \ref{fig:mc_cnd}): 

\begin{itemize}
\item RCS with a condensed surface plus either a condensed (green line) or atmospheric (orange line) hot spot can successfully reproduce the broadband intensity spectrum of 4U 0142+61.
\item RCS with a condensed surface and magnetospheric electron flow with a broken PL momentum distribution (purple line) can closely describe the whole X-ray emission, except in the $7-20$~keV range of the spectrum, where it overpredict the flux.
\item RCS can easily describe the soft X-ray emission in the soft X-ray from 4U~0142+61 (red line), if another mechanism is left responsible for the hard X-ray emission.
\end{itemize}

We do not attempt RCS simulations for a population of hot electrons on top of the condensed surface, as full Comptonization would over-predict the soft X-ray spectrum in the $3-8$ keV range. This is due to the fact that the emission from the condensed surface is intrinsically weakly polarized, with similar surface emission in the O-mode and X-mode; therefore, there are many more O-mode photons available in this case to be Comptonized by hot electrons (in contrast to the low level of O-mode radiation from an atmospheric surface).  This would results in a large flux of O-mode, up-scattered photons. Alternatively, we might be able to get a good description of the soft X-ray considering partial Comptonization (an therefore impose a lower optical depth for the population of hot electrons). Although a detailed treatment of this is left for a future work, we would expect the polarization signature of this process to be similar to the fully Comptonized case where the O-mode dominates in the middle-energy range (5-10~keV).

\begin{table*}
\caption{Input parameters for Monte Carlo RCS code. In all cases the magnetic field at the pole is set to $B=1.25 \times 10^{14}$ G.  The temperatures are given in the frame of the surface, and we assume the radius of the star to be 14~km and its mass to be 1.4~M$_\odot$.}
\label{tab:1}
\begin{tabular}{lllccccccr}
\hline
Surface model &
$T_{\mathrm{pol}}^{\mathrm{eff}}$ &
$T_{\mathrm{hs}}^{\mathrm{eff}}$  &
$\theta_{\mathrm{hs}}$ &
$T_{e}^{\mathrm{c}}$ & 
$\Delta\phi$ &
$e^-$ flow PL &
$\beta_{\mathrm{min}}$ &
$\beta_{\mathrm{break}}$ &
$\gamma_{\mathrm{max}}$ 
\\                
&  [keV] & [keV]               & [deg]                      & [keV] &  [rad] &  &    &   &\\
\hline
\multicolumn{10}{|c|}{4U 0142$+$614}\\
\hline
Atmosphere       & 0.26 & -- & --& -- &  0.4& soft ($\alpha=1.0$)&     0.2& -- & 1.5\\

Atmosphere       & 0.26 & -- & -- & -- &  0.6& broken ($\alpha_1=3.0$ ;
                                                $\alpha_2=1.0$)&         0.2&  0.9& 30.0\\

Atmosphere plus 
hot spot           & 0.26$^a$ & 1.06  & 1.1 & -- &  0.1& hard ($\alpha=1.0$)&    0.9& -- & 30.0 \\

Atmosphere 
plus 
O-mode 
comptonization    & 0.26 & -- & -- & 1.8 &  0.1& hard ($\alpha=1.0$)&     0.9& -- & 30.0\\

\hline

Condensed surface & 0.60 & -- & -- & -- &  0.4&soft ($\alpha=1.0$)&       0.2&  -- & 1.5\\

Condensed surface & 0.60 & -- & -- & -- &  0.6&broken ($\alpha_1=3.0$;
                                            $\alpha_2=1.0$)&              0.2&  0.9& 30.0\\

Condensed surface 
plus condensed 
hot spot          & 0.60$^a$ & 1.4 & 3.0 & -- &  0.1&hard ($\alpha=1.0$)&      0.9&  -- & 30.0\\

Condensed surface 
plus  atmospheric 
hot spot          & 0.60$^a$ & 1.06 & 6.2 & -- &  0.1&hard ($\alpha=1.0$)&       0.9&  -- & 30.0\\

\hline
\multicolumn{10}{|c|}{1RXS J170849.0$-$400910}\\

\hline
Atmosphere plus 
hot spot           & 0.33 & 1.33  & 1.1 & -- &  1.3& hard ($\alpha=1.0$)&    0.3& -- & 30.0 \\
\hline
\multicolumn{10}{|l|}{$^a$ set the thermal map over the whole surface but it is replaced by $T_{\mathrm{hs}}^{\mathrm{eff}}$ at the magnetic pole.}\\
\end{tabular}
\end{table*}

\subsubsection{Condensed surface polarization}
The lower panel in Figure \ref{fig:mc_cnd} shows the polarization signature for the various condensed surface models under RCS. They produce a similar trend above $\sim10$~keV as for the atmospheric models, i.e. the polarization fraction $\sqrt{Q^2+U^2}/I\sim30\%$. This is expected as RCS dominates in the hard X-rays, and we are keeping the same RCS settings (twist angle and momentum distribution for the electron flow) as for the simulations in the atmospheric emission case (see Table \ref{tab:1}). This picture, however, changes substantially in the soft X-rays. A condensed surface with condensed hot spots (green line) produces in general a polarization fraction well below $10\%$ throughout the soft X-ray range. An atmospheric hot spot (orange line), on the other hand, substantially increases the polarization fraction well above $50\%$ in the $5-10$~keV range, while at lower energies the polarization drops as the emission is dominated from the cooler condensed surface.

When we try to reproduce the spectrum of 4U 0142+61 with RCS either in the in $3-10$~keV range (with a single PL momentum distribution, red line), or for both the soft and hard non-thermal emission above $3$~keV  (with a broken PL electron momentum distribution, purple line), the onset of for the RCS $30\%$ polarization fraction moves inside the soft X-rays, in a similar way as in the simulations for the atmospheric surface emission.

\begin{figure}
    \centering
    \includegraphics[width=\columnwidth]{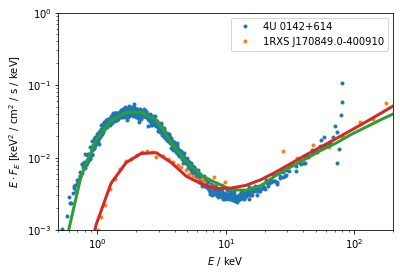}
    \includegraphics[width=\columnwidth]{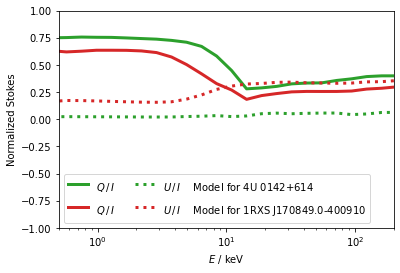}
    \caption{Top panel shows the unfolded Spectral Energy Distributions for 4U~0142+614 from \citet{2015ApJ...808...32T} and 1RXS~J170849.0-400910 from \citet{zane09}. The green and red lines show the RCS model for each source considering  atmospheric emission plus hot spots  (see Table~\ref{tab:1} for surface and RCS parameters). The bottom panel shows the associated Stokes parameters for each RCS model.}
    \label{fig:compare_sources}
\end{figure}

\subsection{1RXS J170849.0-400910}
\label{subsec:1RXS}{}
Although we have focused so far on the source 4U~0142+614, our results also apply to the magnetar 1XRS~J170849.0-400910 with a few modifications.  Observations with RXTE of 1XRS~J170849.0-400910 show pulsations with a period of 11.0~s and spin-down rate $\dot P=1.10 \times 10^{-12}$ \citep{2014ApJ...784...37D}. This translates into a spin-derived magnetic field of $4.7 \times 10^{14}$~G, slightly higher than for 4U~0142+614.  To illustrate the effects of the emission and scattering model, we will hold the magnetic field strength fixed to that of 4U~0142+614 and only vary the parameters of the atmosphere and RCS model to account for the observations of 1XRS~J170849.0-400910.

Fig.~\ref{fig:compare_sources} depicts the spectral energy distributions for the two sources.  The non-thermal component of 1XRS~J170849.0-400910 is much more pronounced relative to the thermal component when compared to 4U~0142+614.  Furthermore, the thermal component in  1XRS~J170849.0-400910  has a slightly higher effective temperature than 4U~0142+614.  \citet{2021arXiv211108584K} fit the spectral energy distribution of 1XRS~J170849.0-400910 with a blackbody or condensed surface thermal component with RCS scattering.  They were not able to fit an atmosphere model to the source, while we find that an atmosphere model with a temperature at the pole of 0.33~keV and with an additional polar hot spot or scattering component does reproduce the spectral energy distribution, so that the analysis for 4U~0142+614 presented earlier can be carried on to this source with a few subtle changes related to the relative contribution of the thermal and non-thermal components.

From Fig~\ref{fig:compare_sources} it is apparent that high-energy scattering is much more important in 1XRS~J170849.0-400910 than it is in 4U~0142+614.  Furthermore, the thermal component is less important than the soft excess, either a hot spot or scattering component, at several keV.   Therefore, we expect that the polarization at low energies will be somewhat lower than in 4U~0142+614, and that the transition to the soft-excess component and high-energy component will occur at somewhat lower energies and more gradually, but, broadly speaking, the polarization signatures present in 4U~0142+614 will be present in 1XRS~J170849.0-400910 as well and will probe the underlying processes in a similar way.

\section{Discussion and conclusions}
\label{sec:discussion_conclusions}

The various emission mechanisms discussed in the previous sections can produce a fairly good description of the broadband X-ray emission of 4U~0142+614. In particular, in the soft X-rays, they can reproduce a similar intensity spectrum. However, the predicted polarization pattern changes substantially between the different mechanisms. Remarkably, IXPE will perform observations of  4U~0142+614 and 1XRS~J170849.0-400910 in the $2-12$ keV range,  a region of the spectrum that can enable us to discriminate between the various mechanisms studied in the this work. 

As found in earlier research work \citep[e.g.][]{gonzalezcaniulef16}, the assumption of a condensed surface reduces the expected polarization to just a few percent in the low-energy range. Since without vacuum birefringence we expect only a small polarization from almost all of the different emission models in the low energy range, this raises the question of what are the prospects to observe the prediction that QED vacuum birefringence preserves the polarization from surface to the observer \citep{heyl00,heyl02}. The only exception is for the atmospheric hotspot model, for which the emission from a few to ten keV would show significant polarization, with or without QED.  At higher energies, the RCS process typically drives the polarization fraction toward 30\%; as both the surface and the scattering layer for RCS \citep{fernandez11} lie well within the polarization limiting radius, we find that even for the condensed surface, the polarized fraction increases to 25\% to 30\% for energies where scattering begins to dominate, while without QED vacuum birefringence, the polarized fraction would remain small even in this regime \citep{taverna20}.  For 4U~0142+614, in the most pessimistic possibility of a fully condensed surface and a hot spot, the transition to the scattering regime that would allow to test the QED effect begins at about 8~keV.  On the other hand, the hard X-ray emission from 1RXS~J170849.0-400910 is stronger relative to the thermal component, so this transition to the scattering dominated regime, even in the case of a condensed surface with a condensed hot spot, begins at about 4~keV.  Consequently, even if the entire surface of both stars are condensed and in both a condensed hot spot accounts for the excess of soft X-rays at a few keV, observations of these sources will be able to test the QED prediction by measuring an increase in the measured polarization through the IXPE band from a few percent at 2~keV to about 25\% at 10~keV due to the contribution of RCS to the polarization.   With these considerations, it is likely that the IXPE observations of 4U~0142+614 and XRS J170849.0-400910 will confirm the QED prediction that a strong magnetic field induces birefrengence in the vacuum. 

As QED outside of this regime is the most precisely tested theory in physics, it is perhaps more interesting to examine how observations of X-ray polarization from magnetars will elucidate their properties.  First, if the emission from magnetars at a few keV indeed comes from their surfaces as currently believed \citep[e.g.][]{1996ApJ...473..322T,1998ApJ...506L..61H}, these observations will determine the state of this surface, if gaseous or condensed.  Despite nearly four decades of observations of magnetars, this will be the first definitive look at the nature of their surfaces. Polarization will also allow us to discriminate between different mechanisms for the non-thermal emission below 10~keV. At higher energies, alternative models beyond those discussed here generally predict that the polarization should be dominated by the ordinary mode even up to high energies \citep[e.g][]{2005ApJ...634..565T,2005MNRAS.362..777H,2019MNRAS.483..599G,thompson20} in contrast to the RCS model, which predicts the dominance of the extraordinary mode. If the surface turns out to be condensed, the lack of polarization at low energies will deprive observers of a lodestone from which to orient the measurements at higher energies to verify the underlying processes for the high-energy emission.  Perhaps with further analysis and more detailed calculations of the alternative models, the direction and extent of polarization as a function of phase and energy could be used to decide among these alternatives and the canonical RCS model, even if the low-energy emission is weakly polarized.

The observations of X-ray polarization from magnetars over the coming years will answer many questions about the emission and processing of radiation from magnetars and will verify the QED prediction of vacuum birefringence.  They are also likely to open and perhaps answer new more detailed questions about magnetars such as what is the temperature distribution on the surface of magnetars (and what does it tell us), how large is the typical field at the surface, what is the distribution of magnetic field in the magnetosphere, what is the strength of the magnetospheric currents and what is their composition (electrons, pairs or electrons and ions).  Measurements of X-ray polarization will indeed open new frontiers on the physics of magnetars, and the models presented here provide a guide for further exploration.

\section*{Acknowledgements}
This work was supported by the Natural Sciences and Engineering Research Council of Canada (NSERC), [funding reference \#CITA 490888-16]. IC is a Sherman Fairchild Fellow at Caltech and thanks the Burke Institute at Caltech for supporting her research. RF acknowledges support from the Natural Sciences and Engineering Research Council of Canada (NSERC) through Discovery Grant RGPIN-2017-04286. This research was enabled in part by support provided Compute Canada (www.computecanada.ca).

\bibliography{main}
\bibliographystyle{mnras}

\label{lastpage}
\end{document}